\documentclass[12pt,a4paper,reqno]{article}

\usepackage{amsmath,latexsym,amssymb,amsthm}
\usepackage{graphicx}
\usepackage{hyperref}
\usepackage{amsmath}\usepackage{amssymb}
\usepackage{amsthm}

\newcommand{\be}{\begin{equation}}
\newcommand{\ee}{\end{equation}}

\newcommand{\B}{\mathcal{B}}

\renewcommand{\S}{\mathcal{S}}

\newcommand{\R}{\mathcal{R}}

\newcommand{\<}{\langle}
\renewcommand{\>}{\rangle}

\newcommand{\tr}{\operatorname{tr}}

\theoremstyle{definition}

\theoremstyle{remark}

\numberwithin{equation}{section}

\begin{document}

\title{\bf{Whose projection postulate?}}

\author{Anthony Sudbery$^1$\\[10pt] \small Department of Mathematics,
University of York, \\[-2pt] \small Heslington, York, England YO10 5DD\\
\small $^1$ tony.sudbery@york.ac.uk}

\date{23 February 2024}

\maketitle

\begin{abstract}

The projection postulate is a description of the effect on a quantum system, assumed to be in a pure state, of a measurement of an observable with a discrete spectrum, in nonrelativistic quantum mechanics. It is often called ``von Neumann's projection postulate'' or ``the L\"uders rule''. This paper is an examination of the versions of this postulate due to Dirac, von Neumann and L\"uders. It is shown that Dirac, in 1930, proposed what is now generally known as the projection postulate. Von Neumann, in 1932, gave a different theory which only applies in special and rather unusual cases.  L\"uders, in 1951, rejected this theory and presented one which is the same as Dirac's. Treatments of observables with continuous spectra by both Dirac and von Neumann are criticised, and the possibility of a generalised version of the projection postulate for this case is considered. The paper concludes with a discussion of the status of the projection postulate (in its various forms) as a separate postulate (independent of the other postulates of quantum mechanics) and as a separate form of time development (in addition to the time-dependent Schr\"odinger equation).


\end{abstract}

\section{Introduction}

The projection postulate of non-relativistic quantum mechanics, in its simplest form (which will be elaborated later in this paper), is the following statement:
\begin{quote}
{\bf P1}\ \  Consider a quantum system $S$ whose pure states belong to a Hilbert space $\mathcal{H}$, and an observable $A$ represented by a self-adjoint operator $\hat{A}$ on $\mathcal{H}$ which has a discrete spectrum. If $A$ is measured  when $S$ is in a pure state $|\psi\>$ and the value $\alpha$ is found, then after the measurement $S$ is in the pure state $\Pi_\alpha|\psi\>$ where $\Pi_\alpha$ is the operator on $\mathcal{H}$ of orthogonal projection onto the subspace of eigenstates of $\hat{A}$ with eigenvalue $\alpha$.
\end{quote}

From the earliest adumbrations of quantum theory by Bohr and Heisenberg, it was recognised that a central feature of the theory was that observation had an inescapable effect on a physical system; after Schr\"odinger's formulation of the theory in terms of wave functions, this came to be known as the ``collapse of the wave function''. The projection postulate makes this vague idea precise. Versions of it were published by Dirac in 1930, by von Neumann in 1932, and by L\"uders in 1951. In the following sections (2, 3 and 4) we will examine each of these and consider their status as ``postulates'' as opposed to \emph{theorems} which can be deduced from a definition of ``measurement'' together with the other postulates of quantum mechanics.

The above statement does not give a full description of the effect of all possible types of measurement. In order to be applicable to all possible measurements in a full physical theory, it needs to be extended to mixed states, to observables with continuous spectra and to measurements whose results have error bars. These extensions will be discussed in Section 5. The penultimate section is a discussion of the status of the projection postulate in the foundations of quantum mechanics, with a brief mention of some of the literature on establishing the validity of the projection postulate and on formulating a version of it in a relativistic theory.

\section{Dirac} 

Paul Dirac's version of the projection postulate is stated on p.\ 49 of the first edition of his book \emph{The Principles of Quantum Mechanics} \cite{Dirac:book}. He bases it on the theorem that any state can be expanded as a linear combination of eigenstates of $\hat{A}$,
\be\label{expansion}
\psi = \sum_\alpha c_\alpha \psi_\alpha
\ee
where $\psi$ and all the eigenvectors $\psi_\alpha$ are normalised (in 1930 Dirac had not yet invented his bra-ket notation). Dirac proposes to use this expansion to obtain ``a mathematical form for the condition that an observation is made with the minimum of disturbance''. Given the expansion, he invites us to ``suppose the observation to be made in such a way that the state of the system afterwards is always one of those that occur in the expansion of the initial $\psi$ in terms of eigen-$\psi$'s of $A$, \emph{i.e.} one of the $\psi_\alpha$'s in \eqref{expansion}''. These states are precisely the projections of $\psi$ onto the eigenspaces of $\hat{A}$, so this ``supposition'' is the projection postulate as stated in the introduction.

To what extent is Dirac proposing an independent postulate?

Dirac takes account of something that is often ignored in more recent discussions of measurements and the projection postulate: there may well be many different ways of measuring a given quantity, and it is likely that they will have different effects on the measured system. It is unreasonable to suppose that there is a single formula for the effect of a measurement of $A$, such as the generally accepted projection postulate. However, one can ask for the least possible disturbance caused by a measurement, and this is what Dirac does. He takes this minimum effect to be given by the projection onto an eigenstate, writing ``This \ldots may conveniently be defined to be the one that causes the minimum of disturbance to the system''. This is indeed postulation. But it need not be; it is clear that, given that the state vector after measurement lies in the eigensubspace with the given eigenvalue, the eigenvector that is closest to the state before measurement, in the metric induced by the inner product of Hilbert space, is just the projection of that state onto the eigenspace. 

There is a further condition that needs to be added to the statement of the projection postulate before it can be regarded as following from the other postulates of quatum mechanics. Dirac argues that the state vector of the system after the measurement of $A$ must be an eigenvector of $\hat{A}$, since it is certain that a repetition of the measurement will give the same result. But this assumes that the system is still present and available to be measured after the first measurement (indeed, this assumption underlies the very idea of a mathematical form for the disturbance to the system). This is not always true; there are measurements which destroy the system, or convert it into a system with a quite different state space -- for example, absorption of a photon by an atom. Measurements which can be repeated in the way required for Dirac's argument to be valid are known as ``measurements of the first kind''. Thus the full statement of the projection postulate should be 
\begin{quote}
{\bf P2}\ \  Consider a quantum system $S$ whose state is the Hilbert space $\mathcal{H}$, and an observable $A$ represented by an self-adjoint operator $\hat{A}$ on $\mathcal{H}$ which has a discrete spectrum. If $A$ is measured by a measurement of the first kind when $S$ is in a pure state $|\psi\>$ and the value $\alpha$ is found, and if this measurement causes the minimum possible change to the state of the system, then after the measurement $S$ is in the pure state $\Pi_\alpha|\psi\>$ where $\Pi_\alpha$ is the operator on $\mathcal{H}$ of orthogonal projection onto the subspace of eigenstates of $\hat{A}$ with eigenvalue $\alpha$.
\end{quote}
As has been pointed out by Isham (\cite{Isham:book} p.\ 35), this is not a postulate but a theorem. In order to establish this, it is necessary to consider the status of the proposition that any state can be expanded as a linear combination of eigenstates. This is indeed a theorem in the mathematical theory of Hilbert and von Neumann if the spectrum of $\hat{A}$ is discrete, but Dirac does not appeal to the mathematical literature; he attempts to prove it himself, but acknowledges that his proof is incomplete. In later editions of his book he simply assumed that this expansion exists, regarding it as a necessary condition for an operator to represent a physical observable. However, a proof meeting Dirac's standard of rigour can be given from the physical postulates of quantum mechanics (\cite{QMPN} p. 51).

Moreover, Dirac does not consider it necessary to restrict his argument to observables with discrete spectra, considering that his invention of the delta function is adequate to deal with continuous spectra. This was trenchantly criticised by von Neumann, as we will see in the next section, and we will respect von Neumann's strictures.


 We note that if the actual result is not examined, the measurement still has a disturbing effect on the system. In this \emph{non-selective} case the post-measurement state must be considered to be the mixed state 
\[
\rho ' = \sum_\alpha P_\alpha|\psi\>\<\psi|P_\alpha .
\]
This need not be considered as a peculiar effect of measurements. It can be derived from a description of the measurement as a physical process of the measured system and the measuring apparatus, proceeding according to the general unitary development of quantum systems: the state of the measured system is a mixed state obtained by tracing over the states of the measuring apparatus. So this version of the projection postulate does not result from assigning any special physical significance to the process of measurement. In general, it can be stated as

\begin{quote} {\bf P3} If the pre-measurement state is a (mixed, in general) state $\rho$, and if the measurement always produces a definite (though unknown) result $\alpha$, then the post-measurement state will be
\be\label{mixed}
\rho ' = \sum_\alpha P_\alpha \rho P_\alpha .
\ee
\end{quote}

\section{Von Neumann}

In 1927 John von Neumann published three papers \cite{vonN:paper1, vonN:paper2, vonN:paper3} on the foundations of quantum mechanics, in reaction to the transformation theory of Jordan, Dirac and Schr\"odinger, whose work depended on ``improper'' functions such as the $\delta$-function whose properties von Neumann regarded as ``absurd''. He therefore set out to give a mathematically rigorous account of this theory, not, as he later wrote (\cite{vonNeumann:QM} p.\ ix), by giving ``a mathematically rigorous refinement and explanation of Dirac's ideas'', but by following ``a procedure differing from the very beginning , namely, the reliance on the Hilbert theory of operators'' . However, these 1927 papers do not contain a precise account of the effects of measurement on a quantum system, but only a statement equivalent to ``collapse of the wave function''. The precise statement can be found in his 1932 book \emph{Mathematical Foundations of Quantum Mechanics} \cite{vonNeumann:QM}. The main burden of this book is to expound the mathematically rigorous theory in the earlier papers, in particular giving a satisfactory treatment of operators with a continuous spectrum. Nevertheless, in his discussion of the effects of measurement von Neumann considers only observables with discrete spectra, arguing that a measurement of an observable with a continuous spectrum cannot give a single number as a result, but can only identify an interval in which the result lies; the measurement is therefore equivalent to a measurement of an observable with a discrete spectrum. If the spectrum is non-degenerate von Neumann argues, in the same way as Dirac, that the state of the system after the measurement must be the eigenstate with the actual result of the measurement as eigenvalue.

If some of the eigenvalues are degenerate, however, von Neumann postulates a different form for the post-measurement state. Suppose the observable $A$ has an eigenvalue $\alpha$ whose eigenspace $\mathcal{S}_\alpha$ has finite dimension {$n_\alpha$ greater than $1$. Von Neumann always considers the general case of mixed states. Let $\rho$ be the statistical operator before the measurement of $A$, $\rho '$ the statistical operator after the measurement; then $\rho '$ can be expanded as
\[ 
\rho ' = \sum_\alpha \sum_i r_{\alpha i}|\psi_{\alpha i}\>\<\psi_{\alpha i}|
\]
where $\{|\psi_{\alpha 1}\>,\ldots |\psi_{\alpha n_\alpha}\>\}$ is an orthonormal basis of $\mathcal{S}_\alpha$, and $0\le r_{\alpha i}\le 1$, while $\sum_{\alpha i}r_{\alpha i} = 1$. Von Neumann assumes that the measurement actually results in one of the states in this basis, so that $r_{\alpha i}$ is the probability that the measurement results in the state $|\psi_{\alpha i}\>$. But he is embarrassed by the fact that this post-measurement state $\rho '$ appears to depend on the choice of the basis $|\psi_{ai}\>$ and thus on more than the pre-measurement state and the quantity $A$ being measured. He therefore looks for cases when $\rho '$ is independent of the choice of basis of $\mathcal{S}_\alpha $, and finds that this is true when $r_{\alpha i} = c_\alpha$ is independent of $i$, so that 
\be\label{vonN}
\rho ' = \sum_\alpha c_\alpha P_\alpha \quad \text{where} \quad c_\alpha = \tr[P_\alpha\rho ']/n_\alpha.
\ee
He declares this to be the post-measurement state in this special case. Thus in this case the measurement leaves the system in one of the eigenspaces $\mathcal{S}_\alpha$, but there is total ignorance as to its position in this subspace. He notes that this requires the pre-measurement state $\rho$ to satisfy 
\[
P_\alpha\rho P_\alpha = c_\alpha P_\alpha
\]
which is ``clearly restricting $\rho$ sharply''; in fact it requires the pre-measurement state to be one of total ignorance as to position in the eigenspace $\mathcal{S}_\alpha$. It is then not surprising that the post-measurement state also exhibits such ignorance. If this condition is not satisfied, von Neumann does not attempt to specify the post-measurement state, saying that ``different arrangements of measurement can actually transform $U$ into different $U'$'' (von Neumann uses $U$ where we use $\rho$). This is the same point that Dirac had made, but von Neumann makes a different choice of assumption to get a definite post-measurement state.

\section{L\"uders}

Gerhart L\"uders criticised von Neumann's assumption that every measurement of a particular observable was associated with a complete orthonormal basis of each eigenspace of the observable, objecting that it led to ``a most complicated mixture'' after the measurement. This is certainly not always warranted; L\"uders gave the extreme example that measurement of the unit operator can be achieved by doing nothing at all. There would then be no disturbance to the measured system, and there is no special orthonormal basis associated with the measurement.

L\"uders proposed to replace von Neumann's prescription by a single statement for each observable, irrespective of the method of measurement. This is the statement {\bf P3}, which follows directly from the proposal of Dirac. Dirac acknowledged that this would not necessarily apply in every actual arrangement for measuring a given observable, but gave a reason (the ``minimum disturbance'') for selecting this process for special consideration. L\"uders, having noted this aspect of von Neumann's discussion of measurement, gave it no further place in his own treatment.

 The projection postulate, as stated in Section 1, is now often called ``the L\"uders rule'',  but  as we have seen, it had already been stated by Dirac. L\"uders claims that it ``shows exactly what is meant by the expression `reduction of  the wave function'{''}, for which the only reference he gives is Pauli's \emph{Handbuch der Physik} treatise \cite{Pauli:book}. We must assume that he was not familiar with Dirac's book.

\section{Continuous spectra}

None of the three writers considered in this paper gave a satisfactory treatment of the measurement of observables with continuous spectra. Dirac regards continuous spectra as being covered by his theory of $\delta$-functions, which he considers makes it possible to treat a continuous spectrum in the same way as a discrete spectrum; he makes only a vague gesture in this direction. Von Neumann states that a measurement of a quantity with a continuous spectrum never yields an exact number as a value of the quantity, but only an interval in which the value lies, a model which he attributes to Wigner. He therefore regards the measurement as equivalent to a measurement of the quantity which specifies in which interval the value lies, a quantity that has a discrete spectrum. L\"uders, in order to ``simplify the treatment'', assumes that all relevant operators have pure point spectra.

It is, of course, true that the statement of the result of a measurement of a continuous quantity is always accompanied by error bars. But von Neumann's assumption that this is a specification of an interval, one of a pre-assigned set of intervals, does not seem to be a universally valid representation of the meaning of error bars. For a start, von Neumann's identification of this as a measurement of an observable whose eigenvalues correspond to set intervals is surely not always correct. It may be appropriate when the error is due to finite instrumental resolution, as for example an instrument with a scale marked in multiples of a given unit; but in other cases the experimenter does not select one of a set of intervals that are given before the experiment. If the result is given as $\alpha\pm\epsilon$, the interval $[\alpha -\epsilon, \alpha + \epsilon]$ has been determined by the experiment, not given beforehand; and it seems wrong to interpret the experimenter's result $\alpha \pm \epsilon$ as a statement that the true value of $A$ certainly lies in the interval $[\alpha - \epsilon, \alpha + \epsilon]$ and not outside it. It is surely much more realistic to regard the stated result $\alpha\pm \epsilon$ as an indication that the value of $A$ is not certain, but is given by a probability distribution with mean $\alpha$ and variance $\epsilon$ (perhaps a normal distribution). This seems to be the only possible interpretation if the result is given, as it often is, in the form $\alpha^{+ \delta}_{- \epsilon}$.

Thus the post-measurement state is a mixed state. The corresponding situation in the case of a discrete spectrum would be an experimental result in the form of a probability distribution $w(\alpha)$ of a discrete variable $\alpha$, and the projection postulate, with a mixed pre-measurement state $\rho$, would give the post-measurement state
\[
\rho' = \sum_\alpha w(\alpha) \frac{P_\alpha \rho P_\alpha}{\tr [P_\alpha \rho P_\alpha]}.
\]
It is not clear what the corresponding expression should be for a general spectrum. However, for a non-selective projection the postulate \eqref{mixed} post-measurement state the postulate {\bf P3} can be re-expressed without explicitly assuming that the observable has a discrete spectrum, as follows.
\begin{quote}
{\bf P4} (conjectural) \  Suppose the observable $A$ is measured when the system is in the state $\rho$. The effect of the measurement is to change the state to $\Pi (\rho)$ where $\Pi$ is the orthogonal projection onto the commutant of $\hat{A}$.
\end{quote}
``Orthogonal'' refers to the Hilbert-Schmidt inner product $(A,B) = \tr (AB^\dagger)$.

A simple calculation shows that if $\hat{A}$ has a discrete spectrum, $\Pi (\rho)$ is a positive operator with trace $1$ if $\rho$ is. In general, however, precise necessary conditions for this, if any, remain to be determined. Unlike the discrete case, this cannot in general be regarded as a theorem resulting from assumptions on the nature of a measurement, in particular repeatability, because an observable capable of repeatable measurement necessarily has a discrete spectrum (\cite{Busch:operational} Chap. IV).

\section{Status of the projection postulate}

We have seen how the projection postulate, in its original form referring to an observable with a discrete spectrum, can be regarded as a theorem stating that the projected post-measurement state necessarily results from certain assumptions about the measurement. This leaves open the question of how this is to be achieved: what physical process will lead to this post-measurement state? A common argument is that it cannot be a unitary evolution, such as is taken to govern physical processes apart from measurements, because unitary evolution cannot take a pure state to a mixed state: the rank of the statistical operator $\rho$ is invariant under unitary transformations $\rho \rightarrow U\rho U^{-1}$. Von Neumann, indeed, asserts that there are two kinds of process in the physical world: Process 1, which takes place when a measurement is made, and Process 2, governed by the time-dependent Schr\"odinger equation, which occurs when the system is not disturbed by measurement. Yet a measurement is a normal process involving physical objects; why should it be governed by different laws from the rest of the physical world? This unsatisfactory situation is the \emph{measurement problem} of nonrelativistic quantum mechanics. How is it to be resolved?

Dirac does not discuss this problem in any edition of his book. John Polkinghorne, in a reminiscence of Dirac, comments ``Physically, Dirac was astonishingly uninterested in the great unresolved interpretative issue of the act of measurement in quantum theory: the collapse of the wave packet is simply attributed to the unanalysed concept of `disturbance'{''}\cite{Polkinghorne:Dirac}. In teaching quantum mechanics, Dirac maintained this attitude; if a student raised the question --- and we did --- he would just brush it aside. This attitude remains common; see, for example, Asher Peres's witty refusal to consider the problem (\cite{Peres:book} p.\ 442).

Von Neumann, on the other hand, took this question seriously. In the final section of his book he constructs a model of the measuring process in which the measuring apparatus is taken to be a quantum system. This leads to a chain of systems that must be considered: the measuring apparatus, for example the mercury column in a thermometer -- light quanta reflected off the mercury column -- the retina of an observer's eye, receiving the light -- the optic nerve of the observer -- their brain. At each stage the quantum state of the whole composite system remains in a pure state, and the projection cannot be understood. Finally, von Neumann considers the consciousness of the observer, correlated with their brain by a principle of psycho-physical parallelism, and finds himself forced to the conclusion that it is the non-physical phenomenon of consciousness that is responsible for the collapse.

However, the contradiction between unitary evolution and the development of a mixed state does not arise if one considers only a subsystem of a whole system which is evolving unitarily. Evolution of the state of a subsystem from a pure state to a mixed state is compatible with unitary evolution of the combined system if the unitary operator entangles the states of the subsystem and the rest of the whole system. Let $\S$ be the state space of the system $S$ which is undergoing measurement, $\R$ the state space of the measuring apparatus and the rest of the universe. If $|\Psi\>$ is a pure state of the whole system, an element of $\S\otimes\R$, then the state of $S$ after the measurement is the mixed state given by the partial trace $\tr_\R|\Psi\>\<\Psi|$, which need not have rank $1$. To pursue von Neumann's analysis into a chain of observing systems, we can take $\S$ to be the whole of this chain up to the brain of the observer, and $\B$ to be the brain of the observer. If one believes in psycho-physical \emph{identity} rather than psycho-physical parallelism, there is no need to follow von Neumann into a non-physical realm of consciousness; we have already reached consciousness. Presumably there are operators representing the contents of consciousness, and a basis of eigenstates of the brain representing possible states of consciousness, including possible observations of the measurement. If one follows the understanding of the quantum state due to Schr\"odinger, Everett and Wheeler (\cite{Schrodinger:interpretns, Everett, Wheeler}), the fact that we always see a single result of a measurement is then a consequence of the fact that each of us can only be aware of --- indeed, can only exist in --- one of these eigenstates of consciousness; which one is seen is governed by the probabilities in the relative state, which must be seen as relative to the observer.

Thus there are three possible answers to the question ``What is the status of the projection postulate?'':

{\bf A1} \ The projection postulate describes a real physical process, peculiar to the context of measurement, in which the Schr\"odinger equation does not apply (von Neumann's Process 1). Instead, a new non-unitary evolution is supposed to occur.

This raises the measurement problem in its most acute form. It has been strongly argued by John Bell \cite{Bell:piddling} that this cannot be accepted: the concept of ``measurement'' is not defined in fundamental physical terms, and cannot have a place in the fundamental laws of physics. 

{\bf A2} \ The projection postulate, in the non-selective form \eqref{mixed}, is a summary of the results of normal unitary evolution, applied to a subsystem of the full system in which the Schr\"odinger equation applies, in a situation in which this unitary evolution converts a separable pure state of the subsystem and the measuring apparatus into an entangled state.

The question of whether and how real measurements do actually conform to a projection postulate in this way has been investigated by many authors, for example Beige and Hegerfeldt \cite{Almut:qjumps}.

This answers Bell's criticism, but it evades the question which is raised by the form {\bf P1} of the projection postulate: it does not address the clear fact that the result of a measurement, in the discrete case, is a single eigenvalue of the measured observable, not a mixture of states with different eigenvalues. There is a third interpretation which meets this objection.

{\bf A3} \ The projection postulate refers to the experience of an observer who is a physical subsystem of the world, and whose experience is described by a one of a set of eigenvectors in the state space of that subsystem. The changing experience of the observer is governed stochastically by the state vector of the universe according to a generalisation of the guidance equation of Bohm and Bell \cite{histories}. If the observer has a memory, the Bell-Bohm evolution results in the memory of an apparent history with repeated projections according to Dirac's postulate, as is demonstrated in a simple model in \cite{histories}. This restores the indeterminism of von Neumann's Process 1, but removes its dependence on the context of measurement. 

It is a matter of perspective whether the universal state vector, or the observer's experience eigenstate, is regarded as a description of ``reality'' \cite{EinsteinTagore}.


Being non-relativistic, the theory considered here, however interpreted, cannot be regarded as a possibly absolutely true theory of physical reality. For theories of measurements in relativistic quantum theory, see \cite{FewsterVerch2}. A difficulty arises in comparing these with the nonrelativistic projection postulate, since quantum field theory is most naturally considered in the Heisenberg picture, while the nonrelativistic projection postulate, at least in the context of von Neumann's Process 1, is most naturally considered in the Schr\"odinger picture and a Heisenberg picture may not be possible. The relation between the relativistic and non-relativistic theories of measurement calls for a separate discussion.  

\section{Conclusion}

 The answer to the question in the title is ``Dirac's''. However, the idea, often associated with the projection postulate, that it describes an unpredictable change in physical reality which happens only when a measurement is performed, interrupting the deterministic evolution described by the Schr\"odinger equation, is due to von Neumann. In this paper we have briefly indicated two ways in which the projection postulate can take a place in an understanding of quantum theory which the state vector is taken to be an objective description of physical reality, evolving according to the time-dependent Schr\"odinger equation without collapsing.

\section*{Acknowledgements}

I am grateful to Fred Muller for making von Neumann's 1927 papers available to me, to Chris Fewster and Eli Hawkins for reading and making valuable comments on earlier drafts of this paper, and to all three of them for helpful discussions.


\begin{thebibliography}{10}

\bibitem{Almut:qjumps}
A.~Beige and G.~C. Hegerfeldt.
\newblock Projection postulate and atomic quantum {Z}eno effect.
\newblock {\em Phys. Rev. A}, 53:53, 1996.
\newblock
  \href{https://arxiv.org/abs/quant-ph/9512012}{arXiv:quant-ph/9512012}.

\bibitem{Bell:piddling}
J.~S. Bell.
\newblock Against ``measurement{"}.
\newblock {\em Physics World}, 3(8):33--40, August 1990.

\bibitem{Busch:operational}
Paul Busch, Marian Grabowski, and Pekka~J. Lahti.
\newblock {\em Operational quantum physics}.
\newblock Springer, 1997.

\bibitem{Dirac:book}
P.~A.~M. Dirac.
\newblock {\em The Principles of Quantum Mechanics}.
\newblock Oxford University Press, 1930.

\bibitem{FewsterVerch2}
Christopher~J. Fewster and Rainer Verch.
\newblock Measurement in quantum field theory.
\newblock {\em Encyclopedia of Mathematical Physics}, to appear.
\newblock \href{https://arxiv.org/abs/2304.13356}{arXiv:2304.13356}.

\bibitem{Everett}
Hugh~Everett III.
\newblock ``{R}elative state'' formulation of quantum mechanics.
\newblock {\em Rev. Mod. Phys.}, 29:141--153, 1957.

\bibitem{Isham:book}
C.~J. Isham.
\newblock {\em Lectures on Quantum Theory}.
\newblock Imperial College Press, 1995.

\bibitem{Pauli:book}
Wolfgang Pauli.
\newblock {\em General Principles of Quantum Mechanics}.
\newblock Springer-Verlag, 1980.

\bibitem{Peres:book}
Asher Peres.
\newblock {\em Quantum Theory: Concepts and Methods}.
\newblock Kluwer, 1993.

\bibitem{Polkinghorne:Dirac}
J.~C. Polkinghorne.
\newblock Dirac and the interpretation of quantum mechanics.
\newblock In Behram~N. Kursunoglu and Eugene~P. Wigner, editors, {\em {P}aul
  {A}drien {M}aurice {D}irac: reminiscences about a great physicist}, pages
  228--230. Cambridge University Press, 1987.

\bibitem{Schrodinger:interpretns}
Erwin Schr{\"o}dinger.
\newblock July 1952 colloquium.
\newblock In M.~Bitbol, editor, {\em The Interpretation of Quantum Mechanics:
  Dublin Seminars (1949-1955) and other unpublished essays}, pages 19--37. Ox
  Bow Press, 1996.

\bibitem{QMPN}
Anthony Sudbery.
\newblock {\em Quantum Mechanics and the Particles of Nature}.
\newblock Cambridge University Press, 1986.

\bibitem{EinsteinTagore}
Anthony Sudbery.
\newblock Einstein and {T}agore, {N}ewton and {B}lake, {E}verett and {B}ohr:
  the dual nature of reality.
\newblock In P.~Ghose, editor, {\em The nature of reality: the perennial
  debate}. Routledge, 2016.
\newblock {\tt arXiv:1205.1479}.

\bibitem{histories}
Anthony Sudbery.
\newblock Histories without collapse.
\newblock {\em Int. J. Theor. Phys.}, 61:39, 2022.
\newblock \href{https://arxiv.org/abs/2012.13430}{arXiv:2012.13430}.

\bibitem{contsreduction}
Tony Sudbery.
\newblock Continuous state reduction.
\newblock In R.~Penrose and C.~J. Isham, editors, {\em Quantum concepts in
  space and time}, pages 65--83. Oxford University Press, 1986.

\bibitem{vonN:paper1}
J.~von Neumann.
\newblock Mathematische {B}egr{\"u}ndung der {Q}uantenmechanik.
\newblock {\em Nachrichten von der Gesellschaft der Wissenschaften zu
  G{\"o}ttingen}, pages 1--57, 1927.

\bibitem{vonN:paper3}
J.~von Neumann.
\newblock Thermodynamik quantenmechanischer {G}esamtheiten.
\newblock {\em Nachrichten von der {G}esellschaft der {W}issenschaften zu
  {G}{\"o}ttingen}, pages 273--291, 1927.

\bibitem{vonN:paper2}
J.~von Neumann.
\newblock Wahrscheinlichkeitstheoretischer {A}ufbau der {Q}uantenmechanik.
\newblock {\em Nachrichten von der {G}esellschaft der {W}issenschaften zu
  {G}{\"o}ttingen}, pages 245--272, 1927.

\bibitem{vonNeumann:QM}
J.~von Neumann.
\newblock {\em Mathematical Foundations of Quantum Mechanics}.
\newblock Princeton University Press, 1955.
\newblock translated by Robert T. Beyer. Originally published 1932.

\bibitem{Wheeler}
J.~A. Wheeler.
\newblock Assessment of {E}verett's ``relative state'' formulation of quantum
  theory.
\newblock {\em Rev. Mod. Phys.}, 29:463--465, 1957.

\end{thebibliography}

\end{document}